\documentclass[conference]{IEEEtran}
\usepackage{cite}
\usepackage{graphicx}

\graphicspath{ {images/} }
\IEEEoverridecommandlockouts
\usepackage{cite}
\usepackage{amsmath,amssymb,amsfonts,bbm}
\usepackage{subfigure,bm,balance}
\usepackage[ruled,vlined]{algorithm2e}
\usepackage{graphicx}
\usepackage{textcomp}
\usepackage{xcolor}

\begin{document}


\title{NUMSnet: Nested-U Multi-class Segmentation network for 3D Medical Image Stacks}
\author{Sohini Roychowdhury,~\IEEEmembership{Member,~IEEE}\\
\IEEEauthorblockA{Adjunct Faculty, Santa Clara University,\\
Accenture LLP, USA,\\
Email: roych@uw.edu}
\vspace{-0.5cm}
}

\maketitle
\begin{abstract}
Semantic segmentation for medical 3D image stacks enables accurate volumetric reconstructions, computer-aided diagnostics and follow up treatment planning. In this work, we present a novel variant of the Unet model called the NUMSnet that transmits pixel neighborhood features across scans through nested layers to achieve accurate multi-class semantic segmentations with minimal training data. We analyze the semantic segmentation performance of the NUMSnet model in comparison with several Unet model variants to segment 3-7 regions of interest using only 10\% of images for training per Lung-CT and Heart-CT volumetric image stacks. The proposed NUMSnet model achieves up to 20\% improvement in segmentation recall with 4-9\% improvement in Dice scores for Lung-CT stacks and 2.5-10\% improvement in Dice scores for Heart-CT stacks when compared to the Unet++ model. The NUMSnet model needs to be trained by ordered images around the central scan of each volumetric stack. Propagation of image feature information from the 6 nested layers of the Unet++ model are found to have better computation and segmentation performances than propagation of all up-sampling layers in a Unet++ model. The NUMSnet model achieves comparable segmentation performances to existing works, while being trained on as low as 5\% of the training images. Also, transfer learning allows faster convergence of the NUMSnet model for multi-class semantic segmentation from pathology in Lung-CT images to cardiac segmentations in Heart-CT stacks. Thus, the proposed model can standardize multi-class semantic segmentation on a variety of volumetric image stacks with minimal training dataset. This can significantly reduce the cost, time and inter-observer variabilities associated with computer-aided detections and treatment. 
\end{abstract}


\section{Introduction}

Multi-class semantic segmentation of regions of interest (ROIs) from medical 3D image stacks of CT or MRI images is important for diagnostic pathogenesis and for pre-procedural planning tasks. Performing such segmentations manually can be both costly and time intensive \cite{intro}. Additionally, manual segmentation process suffers from inter-observer variabilities where two medical practitioners may disagree between the exact locations of the ROIs \cite{intro}. In such situations, traditional deep learning approaches for segmentation have been used largely to augment the human effort needed to isolate the ROIs \cite{unetsegnet}. The idea is for a limited number of images to be manually annotated, followed by training a deep learning model on the hand annotated data, and generating standardized segmentations across all future image frames. The challenge here is that most deep learning approaches are data hungry and require large volumes of initial annotated data to yield standardized ROIs, especially if there are many ROIs, or ROIs with varying sizes. Also, medical 3D image stacks often represent variable pixel resolution and noise across imaging equipment’s, which impedes extendability of the automated deep learning solutions to other image stacks. Several existing works for medical image semantic segmentation perform binary segmentations per image \cite{unet}\cite{unetp}, or two-stage multi-class segmentations for image stacks \cite{unetp2}. In this work, we present a novel single-stage variant of the popular Unet model that contains additional nested layers to capture the spatial neighborhood characteristics and propagate it across image scans for accurate multi-class semantic segmentation performances with a minimal training set of images.

\begin{figure*}[ht!]
    \centering
    \includegraphics[width=4.5in]{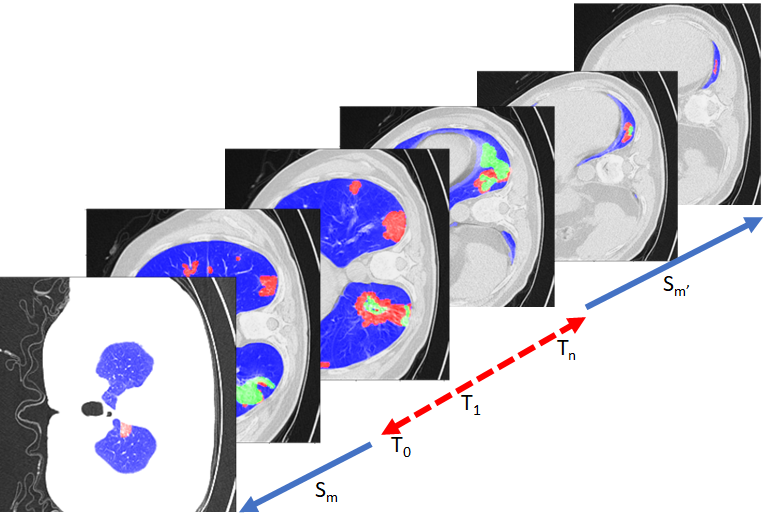}
    \caption{An example of the proposed NUMSnet system on a Lung-CT image stack. The training images ($T$) are selected in order or in sequence while test images ($S$) can be random or in sequence.}
    \label{system}
\end{figure*}

Unet models and their variants have been the preferred deep learning model in medical image processing and detection domains due to the low computational complexity. This allows training from a few hundred images as opposed to thousands of annotated images typically needed for other imaging domains such as autonomous drive and augmented reality \cite{intro2} \cite{intro3}. Thus, the lack of high volumes of quality annotated data, and inter-observer variability \cite{interobserver} makes Unet and its variant models the preferable method for ROI segmentation and computer aided detection. The proposed Unet variant model, called the NUMSnet, propagates image features across scans, which results in faster network convergence with few training images for volumetric medical image stacks. The NUMSnet requires training images in order, but not necessarily subsequent images in a sequence. The training set is shown in Fig. \ref{system} by the image stack [$T_0$ to $T_n$]. Once trained, the test set of images can be ordered or be randomized per stack as represented by sets $S_m$ and $S_{m'}$ in in Fig. \ref{system}.

In this paper we present a novel multi-class semantic segmentation model that propagates layers across consecutive scans to achieve multi-class semantic segmentation with only 10\% of frames per 3D image stack. We investigate three main analytical questions towards multi-class semantic segmentation in 3D medical image stacks. 1) Does transmission of image features from some of the layers of a Unet variant model enhance semantic segmentation performance for multi-class segmentation tasks? 2) Is the order of training and test frames significant to segmentation tasks for 3D volumes? 3) How many layers should be optimally propagated to ensure model optimality while working with sparse training data? The key contributions in this work are as follows.
\begin{enumerate}
    \item A novel multi scan semantic segmentation model that propagates feature-level information from a few nested layers across ordered scans to enable feature learning from as few as 10\% annotated images per 3D medical image stack.
    \item Transfer learning performance analysis of the proposed model with respect to existing Unet variants on multiple CT image stacks from the Lung-CT (thoracic region) scans to Heart-CT regions. The NUMSnet model achieves up to 20\% improvement in segmentation recall and 4-9\% improvement in Dice scores for multi-class semantic segmentations across image stacks.
    \item Identification of optimally located minimal training images per volumetric stack for multi-class semantic segmentation.
    \item Identification of the optimal number of layers that can be transmitted across scans to prevent model over-fitting for segmentation of up to 7 ROIs with variables shapes and sizes.
\end{enumerate}

This paper is organized as follows. Existing literature and related works are reviewed in Section \ref{rel}. The datasets under analysis and the NUMSnet model are explained in Section \ref{methods}. The experiments and results are shown in Section \ref{exp}. The conclusions are in Section \ref{conclusion} and relevant discussions and limiting conditions are presented in Section \ref{disc}.

\section{Related work}\label{rel}
Deep learning models have been heavily popular for computer aided detections in the past decade over signal processing methods in \cite{signal1}\cite{signal2}. This is primarily due to the ability of deep learning models to automatically learn features that are indicative of a ROI if a significant volume of annotated data is provided. Signal processing models on the other hand rely on hand generated features that may lead to faulty detections due to the high variabilities across imaging modalities, storage and transmission formats. The prior work in \cite{lungseg} demonstrates a two path CNN model that can take filtered Lung-CT images followed by fuzzy c-means clustering to segment the opacity in per Lung-CT image. While such feature-based works have low data dependence, the models often do not scale across datasets.

Unet models have been used significantly for medical image segmentation applications since 2015 \cite{unet}. While other deep learning models such as MaskRCNN \cite{maskrcnn} and Fully Convolutional Neural networks (FCNs) \cite{FCN} are more popular in non-medical domains, Unet and its variants have continued to be the most preferred deep learning model for medical image segmentation tasks. Unet and its variant models apply long and short skip connections that ensure the number of trainable parameters are low, thereby leading to lower computational complexity and better ease of train-ability with fewer training images. Other variants of U-net models include 3D models such as V-net \cite{vnet} and \cite{unetp2} that apply the Unet model connections at a voxel/volume level. These methods incur high computational complexity and require several sequential scans to be labelled for training.

Over the last few years, several Unet model variants have been applied for dense volumetric scan segmentations. In instances where high volumes of annotated data are readily available, such as anatomical regions in heart-CT scans in \cite{heartct} \cite{unetp2}, multi-stage Unet variants have introduced. The works in \cite {payer} and \cite{unetp2} train two separate Unet models with separate loss functions with the objective of zooming into the foreground regions in the first network followed by separating the foreground in to the various ROIs. Another work in \cite{heartct} implements a deeply supervised 3D Unet model with multi-branch residual network and deep feature fusion along with focal loss to achieve improved segmentations for smaller ROIs. Other variants of multi-Unet models such as the work in \cite{aymar} implements trained Unet models at different resolutions, i.e., one Unet model trained on images of dimensions [256x256], while another trained at resolution [512x512] and so on for lung segmentation. However, these methods require significantly high volumes of annotated data to train the multiple Unet models or to perform 3D volumetric convolutions. 

Other recent works in \cite{unetsegnet} and \cite{covidnet} have applied variations to the Unet model to achieve segmentation of opacity and lung regions from chest-CT scans to aid COVID-19 detections. Also, in \cite{infnet}, an Inf-net and Semi-Inf net models are presented that can achieve binary segmentation performances for lung opacity detection with Dice scores in the range of 0.74-0.76. Most of these existing methods require several hundred annotated training images across scans and patients and can efficiently be trained for binary semantic segmentation tasks. In this work, we present a novel Unet model variant that propagates deep learning layer information across scans, thereby achieving superior multi-class semantic segmentation performances to most existing methods while being trained by fewer than a hundred annotated scans overall.

Some of the well known Unet model variants used in the medical imaging domain are the wide Unet (wU-net) and Nested Unet (Unet++) \cite{unetp}. While a typical Unet model of depth 5 will has filter kernel widths [32,64,128,256,512] at model depth 1 through 5, the wUnet model has filter kernel widths [35,70,140,280,560] at model depth 1 through 5. Thus, the wUnet has more parameters, and thereby can enhance segmentation performances when compared to Unet. The Unet++ model on the other hand generates dense connections with nested up-sampled layers to further enhance the performances of semantic segmentation as presented in \cite{unetp3} \cite{unetp4}. In this work, we propose an enhanced Unet++ architecture called the NUMSnet, where the features from the nested up-sampled layers are transmitted across scans for increased attention to smaller regions of interest (such as opacity in Lung-CT images). This layer propagation across scans enables multi-class semantic segmentation with only 10\% of annotated images per 3D volume stack.

\section{Materials and Methods}\label{methods}
\subsection{Data: Lung-CT and Heart-CT Stacks}
In this work we analyze two kinds of single plane volumetric CT image stacks. The first category of Lung-CT image stacks are collected from the Italian Society of Medical and Interventional Radiology. The first Lung-CT (Lung-med) volumetric stack \cite{covidct} contains 829 images from a single 3D image stack with [512x512] dimension images. 373 out of these 829 scans are annotated. The second dataset (Lung-rad) contains 9 axial volume chest CT scans with 39-418 images per stack. All Lung-CT images are annotated for 3 ROIs namely: ground-glass opacity (GGO), consolidations and the Lung region as foreground and can be downloaded from \cite{lungdata}. 

\begin{figure*}[ht!]
    \centering
    \includegraphics[height=4.5in]{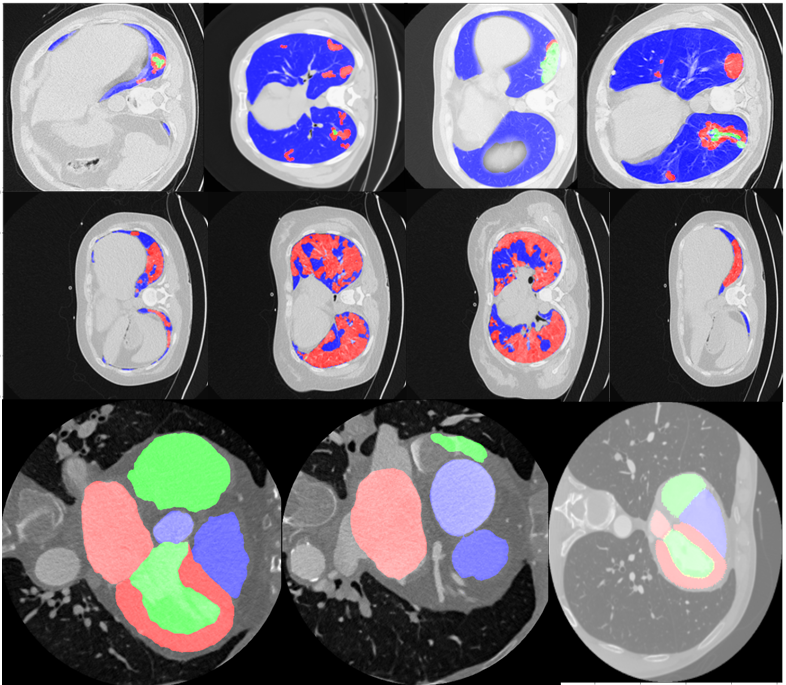}
    \caption{Examples of multi-class segmentation datasets used in this work. Row 1: Lung-med dataset, Row 2: Lung-rad dataset. For Row 1 and Row 2 the regional color coding is as follows. Blue: Lung region, Red: GGO, Green: consolidation regions. Row 3: Heart-CT dataset. The ROIs are color coded as follows. Red plane: label pixels 205 and 420. Blue plane: label pixels 500 and 550. Green plane: label pixels 600, 820 and 850.}
    \label{fig:res1}
\end{figure*}

The second category of Heart-CT image dataset is from the MICCAI 2017 Multi-Modality Whole Heart Segmentation (MM-WHS) challenge \cite{unetp2}\cite{heartct} from where we select the first 10 training CT image stacks of the heart region for analysis. This dataset contains coronal volumetric stacks with 116-358 images per volume and multi-class semantic segmentation annotations for up to 7 heart-specific ROIs represented by label pixel values [205, 420, 500, 550, 600, 820, 850], respectively. These pixel regions represent the [left ventricle blood cavity (LV), myocardium of the left ventricle (Myo), right ventricle blood cavity (RV), left atrium blood cavity (LA), right atrium blood cavity (RA), ascending aorta (AA) and the pulmonary artery (PA)], respectively. It is noteworthy that for the Heart-CT dataset only 10-15\% of the images per stack contain the annotated ROIs. Thus, at the time of ordered training dataset selection, it is ensured that atleast 50\% of the training samples contain annotations. Some examples of the Lung-CT and Heart-CT images and their respective annotations are shown in Fig. \ref{fig:res1}.

Each image from the data stacks under analysis here is pre-processed for the Unet and variant models. First, each input image is resized to [256x256x1] for ease of processing. Next, the resized image $I$ is re-scaled to the range [0,1], thereby resulting in image $I'$, using min-max normalization as shown in \eqref{minmax}, where, $min_I$ and $max_I$ refer to the minimum and maximum pixel values in $I$. This is followed by generation of multi-dimensional label vectors [256x256x$d$] per image, where $d$ represents the number of classes that each pixel can be classified into. These label vectors are generated as a binary images per class. For example, the Heart-CT stack images contain up to 7 different annotated regions depicted by a certain pixel value $pix_i, \forall i=[1:7]$. Thus, the ground-truth label vector ($G'$) generated per image contains 7 planes, where each plane $G'_i$ is generated as a binary mask from the label masks ($G$) as shown in (2). This process defines the ground-truth G' such that the Unet decision making function ($f_i$) proceeds to analyze if each pixel belongs to a particular class $i$ or not. Finally, the output is a $d$ dimensional binary image ($P$) where each image plane ($P_i$) is thresholded at pixel value $\tau=0.5$ as shown in (3).

\begin{align}\label{minmax}
    I'=\frac{I-min_I}{max_I-min_I}.\\
    \forall i \in [1:d], G'_i=[G==pix_{i}],\\
    and, P_i=[f_i(I')>\tau].
\end{align}

Once the datasets are pre-processed, the next step is to separate the data stacks to training, validation and test sets. There are two ways in which the training/validation/test data sets are sampled per volume stack. For the first, random sampling method, 10\% of the scans per volume are randomly selected in ascending order as training, 1\% of the remaining images are randomly selected for validation and all remaining images are used for testing. The second sequential sampling method starts from a reference scan in the volumetric stack. This reference scan could either be the first or the middle scan in the stack. We sample 10\% of the total number of images in the stack starting from the reference scan in sequence and these become the training set of images. From the remaining images, 1\% can be randomly selected for validation, while all remaining scans are test set images in sequence. Using these methods, we generate training sets of size: [82x256x256x1],[84x256x256x1] and [363x256x256x1], respectively, for the Lung-med, Lung-rad and Heart-CT stacks, respectively.

\subsection{Unet Model and Variants}
\begin{figure*}[ht!]
    \centering
    \includegraphics[width=5.0in, height=2.5in]{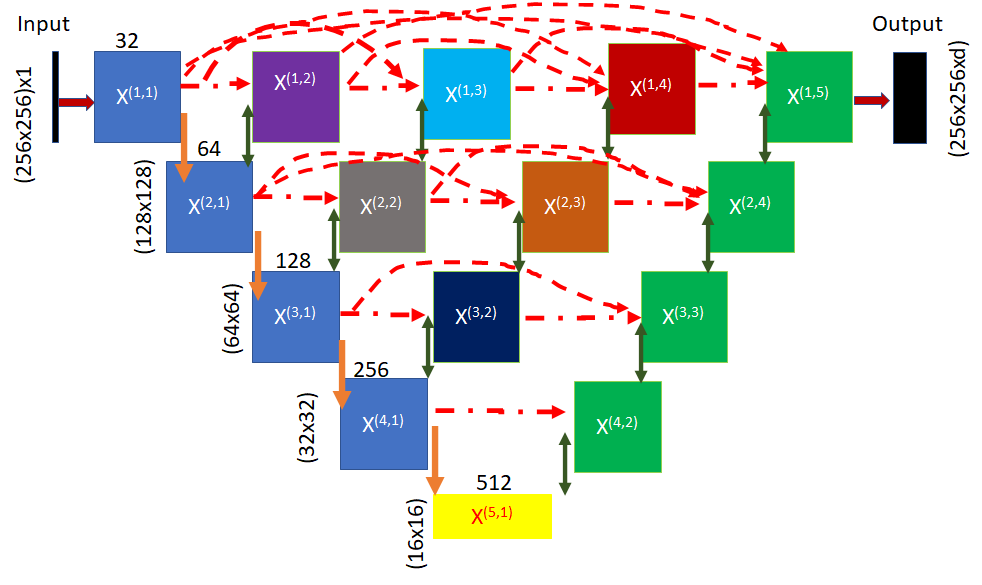}
    \caption{Example of a Unet++ model for depth 5. The blue layers correspond to convolved and pooled layers. The green layers correspond to merged transposed convolutions followed by convolution outcomes from the same depth layers. The 6 additional nested color coded layers as (purple, cyan, red, grey, orange, dark blue) corresponding to [X(1,2), X(1,3) X(1,4) X(2,2) X(2,3) X(3,2)], respectively, contain spatial pixel neighborhood information that can be transmitted temporally across images/scans for increased accuracy of semantic segmentation.}
    \label{fig:unet}
\end{figure*}

Till date, Unet-variants such as wUnet, Vnet and Unet++ models, have been applied to improve foreground segmentation precision for small regions of interest as shown in \cite{unetp}\cite{heartct}. It is noteworthy that for binary segmentation tasks, the relative variation in performances for such Unet model variants remains less significant \cite{girish}. However, to improve multi-class semantic segmentations, we propose a variant of the Unet++ or the Nested-Unet model from \cite{unetp}. One major difference between the Unet++ and the traditional Unet model is the presence of nested layers that combine the convolved and pooled layers with the up-sampled transposed convolutional layers at the same level. Thus, for a Unet with depth of 5, a Unet++ model results in 6 additional nested layers shown as [X(1,2), X(1,3) X(1,4) X(2,2) X(2,3) X(3,2)] in Fig \ref{fig:unet}. These additional layers increase signal strengths at each depth level and amplify the segmentation decisions around boundary regions of ROIs \cite{unetp}.

The process of semantic segmentation using Unet for a single plain gray-scale medical image proceeds as follows. The input image $I'$ is subjected to 2D convolutions and pooling in layers [X(1,1), X(2,1), X(3,1), X(4,1)], respectively. The output of each layer results in an image with half the input dimensions but additional feature planes. For example, the input to layer X(1,1) is the image of size [256x256x1], while the output has dimensions [128x128x32], due to convolution with a [3x3] kernel with width 32 and max-pooling with a [2x2] kernel. Thus, at the end of the fifth layer (X(1,5)), a feature vector of size [16x16x512] is generated. At this point, the transposed convolutions in 2D are performed with kernel size [2x2] to up-sample the input images from the previous layer. One key consideration for all the up-sampling layers is that to promote better distinction between foreground pixels (scaled value 1) versus background pixels (scaled value 0), the images/features from same depth are concatenated followed by the transposed convolutions. For example, at depth 4, the output from layer X(4,1) is concatenated with the up-sampled image from layer (1,5), resulting in image features of dimension [32x32x512] that are then subjected to convolutions in layer (4,2), thereby resulting in image features of dimension [32x32x256].This up-sampling, concatenation and convolution process continues till the output of layer X(1,5) is an image ($P$) with dimensions [256x256x$d$], $d$ being the number of planes corresponding to ROIs.

The Unet++ model, on the other hand, was developed to enhance the boundary regions for relatively small ROIs by introducing nested up-sampling layers at each depth level as shown in Fig. \ref{fig:unet}. For example, the input for layer X(1,2) has the size of [256x256x32] being an output from the X(1,1) layer, which is then concatenated with the transposed convolved output of layer X(2,1) with same dimensions. The layer X(1,2) then performs convolutions to generate [256x256x32] image feature as output that then feeds into the layer X(1,3). This process continues across all the nested up-sampling layers. 

The primary parameters that need to be tuned to ensure optimally trained Unet or a variant model are the following: data augmentation methods, batch size, loss function, learning rate and reported metric per epoch. In this work, we apply image data augmentation using the tensorflow keras library by augmenting images randomly to ensure rotation range, width shift range, height shift range and shear range of 0.2, respectively, and zoom range of [0.8,1] per image. Since the training data set has few samples, we implement a training batch size of 5 for the Lung-CT images and batch size of 10 for heart CT images. It is noteworthy that batch size should scale with the number of detection classes, thus, we use additional images per batch for the Heart-CT stack. For all the Unet and variant models we use the Adam optimizer with learning rate of $10^{-3}$. Finally, the metrics under analysis are shown in \eqref{dmetric}-(7) based on the work in \cite{metrics}. For each image with $l$ pixels and $d$ images planes for the ground-truth ($G'_i$), intersection over union ($IoU$) or Jaccard metric in (4) represents the average fraction of correctly identified ROI pixels. The Dice coefficient $Dice$ in (5) further amplifies the fraction of correctly classified foreground pixels. Precision ($Pr$) in (6) and recall ($Re$) in (7) denote the average fraction of correctly detected ROI pixels per predicted image and per ground-truth image plane, respectively.
\begin{align}\label{dmetric}
IoU=\sum_{i=1}^{d}\sum_{j=1}^{l}\frac{|P_i(j)\cap G'_i(j)|}{P_i\cup G'_i},\\
Dice=\sum_{i=1}^{d}\sum_{j=1}^{l}\frac{2*|P_i(j)\cap G'_i(j)+1|}{P_i(j)+G'_i(j)+1},\\
Pr= \sum_{i=1}^{d}\sum_{j=1}^{l}\frac{P_i(j)\cap G'_i(j)}{P_i(j)},\\
Re= \sum_{i=1}^{d}\sum_{j=1}^{l}\frac{P_i(j)\cap G'_i(j)}{G'_i(j)}.
\end{align}

The loss functions under analysis are shown in \eqref{dloss}-(10). The Dice coefficient loss ($DL$) in (8) is inverse of the Dice coefficient, so it ensures the average fraction of correctly detected foreground regions increases per epoch. The binary cross entropy loss ($BCL$) in (9) is a standard entropy based measure that decreases as the predictions and ground-truth become more alike. Finally, the binary cross entropy-Dice loss ($BDL$) in (10) is a combination of BCL and DL based on the work in \cite{unetp}. 
\begin{align}\label{dloss}
DL=-D,\\
BCL=-\sum_{i=1}^{d}\sum_{j=1}^{l}[P_i(j)log(G'_i(j))],\\
BDL=\frac{BCL}{2}+DL.
\end{align}

\begin{figure*}[ht!]
    \centering
    \includegraphics[width=4.0in]{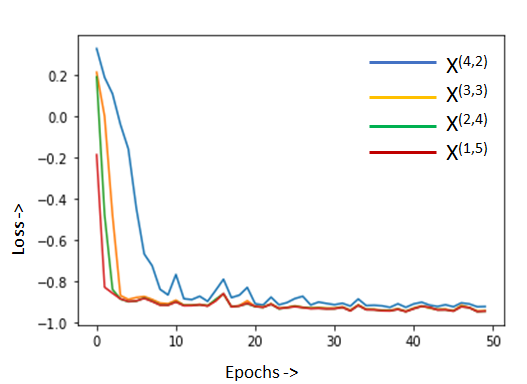}
    \caption{Example of loss functions per depth layer in Unet++ model using the deep-supervision feature on the Lung-med training dataset. The resized image outcome from X(4,2) achieves lower segmentation resolutions when compared to the outcome from X(1,5). Thus, nested layers enhance local boundary-region specific features for segmentation.}
    \label{fig:loss}
\end{figure*}

Finally, we analyze the loss function curves per epoch using the deep-supervision feature from the Unet++ model \cite{unetp} in Fig. \ref{fig:loss}. Here, we assess convergence rates for outputs at each depth levels. From Fig. \ref{fig:loss}, we observe that the convergence of outputs from depth 1 and 2 (layers X(1,5) and resized output of X(2,4)) are relatively similar and better than the loss curves for depth 4 (resized output of layer X(4,2)). This implies that as the transposed convolutions move further away from the dense feature layer X(5,1), additional local feature-level information gets added to the semantic segmentation output. Thus, for a \textit{well-trained} Unet++ model, the initial transposed convolution layers closer to the global feature layer X(5,1) bring less value to the semantic segmentation task when compared to the farther away layers from X(5,1), i.e. layers X(1,2),X(1,3),X(1,4). This variation in loss curves at the different depth levels, based on the work in \cite{qunet} demonstrates the importance of the additional up-sampling nested layers towards the final multi-class segmented image.

\subsection{The NUMSnet Model}
In the recent years Unet models and its variants have been trained and implemented for specific binary segmentation tasks such as COVID screening using Lung-CT images \cite{covidct}. While the Unet and variant models are efficient at segmentations per scan, segmenting volume stacks need further intervention wherein pixel neighborhood information can be transmitted to the next ordered scan, thereby allowing better resolution of semantic segmentations by training on a few images. The NUMSnet model is a 3D extension to the Unet++ model, wherein the outcomes of the nested Unet++ layers from each image are transmitted to the next scan. For each subsequent scan, the transmitted layer image features from the previous scan are concatenated and convolved with the same layer equivalent of the current image features followed by the regular convolution pooling and up sampling operations of the remaining layers as explained in the previous subsection. For example, the output from layer X$^{n}$(1,2) from the $n$-th training image with dimensions [256x256x32] is concatenated with the output of layer X$^{n+1}$(1,2) from the very next training image and convolved with a [3x3] kernel to result in [245x256x32] dimension image/features output for the layer X(1,2) for the training image $n+1$. Similarly the outputs of the other nested layers X(1,3), X(1,4), X(2,2), X(2,3), X(3,2) are propagated to the next ordered scan. For an optimal NUMSnet model, we apply batch normalization to encoder layers only and dropout at layers X(4,1),X(5,1) only. \footnote{Github Code available at https://github.com/sohiniroych/NUMSnet} Also, the widths of kernels per depth layer for the NUMSnet model are [5,70,140,280,56] similar to that of the wUnet model. This process of transmitting and concatenating layer-specific features with those of the subsequent ordered images generate finer boundary condition outcomes. This variation in the Unet++ model to generate the NUMSnet model is shown in Fig. \ref{fig:model}. The additional layers generated in this process are shown in the model diagrams in the Appendix section Fig. \ref{fig:numsnet}.
\begin{figure*}[ht!]
    \centering
    \includegraphics[width=3.3in, height=2.5in]{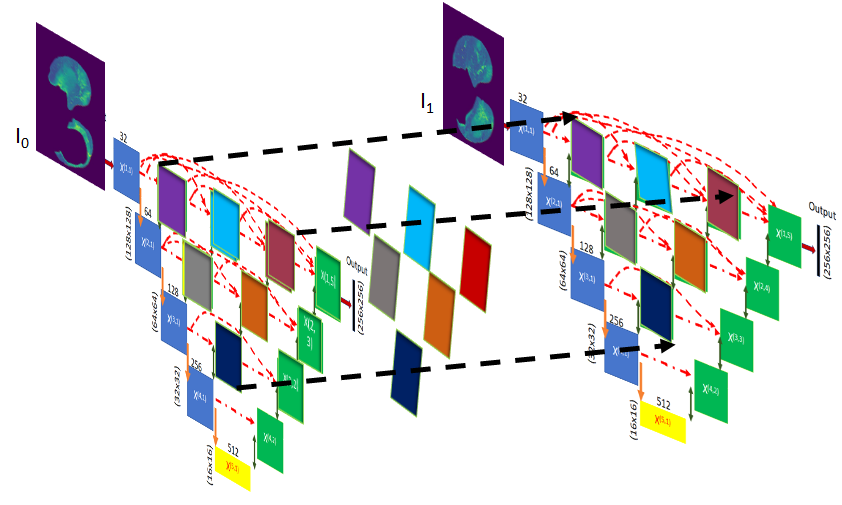}
    \caption{The proposed NUMSnet that propagates the image features from the 6 nested layers across scans. Outcome of each nested layer is concatenated and convolved with the equivalent layer of the subsequent ordered image in the 3D stack.}
    \label{fig:model}
 \end{figure*}

The NUMsnet model processes the 3D medical image stacks as follows. For the first image in the training stack, the nested layer outcomes are convolved with themselves due to the lack of a previous layer image features. Next, as the training progresses, the 6 nested layer outputs and the model layer states are collected for each scan and propagated to the next scan. The weights and biases from model neurons back-propagate to minimize the loss function and this process continues per training batch and epoch. It is noteworthy that while the training samples are ordered, the test samples may be out of order starting at the other end of the stack or starting at a new volumetric stack. In the testing phase, the nested layer outputs and model layer weights and biases are collected per test image and passed to the next image. Once the NUMSnet model is optimally trained, the out of order scans in test stacks do not significantly impact the segmentation outcomes. All other parameters including data augmentation, loss functions, batch size, compiler, learning rate and reported metrics are kept similar to that of the Unet model and variants to realize the segmentation enhancements per epoch.

The primary improvement in semantic segmentation capability for a volume stack introduced by NUMsnet over a Unet model \cite{unet} and Unet++ \cite{unetp} model is the additional skip connections across scans that magnify pixel neighborhood features across scans. For medical images, the relative variation in pixel neighborhoods is significantly lesser than regular camera acquired images like those for autonomous driving or satellite imagery \cite{intro2}. Thus, the feature-level propagation across scans enhances the decision making around boundary regions especially for smaller ROIs. However, the additional nested layer concatenation introduces higher number of parameters in the Unet-variant models, which leads to slower training time and higher GPU memory requirements for model training. In this work, we use Nvidia RTX 3070 with 8GB of GPU RAM on an Ubuntu Laptop and tensorflow/keras libraries to train and test the volume segmentation performances. In instances where models have a high number of parameters, keeping a small batch size of 5-10 ensures optimal model training. As an estimate for the model computational complexity, the number of trainable parameters in Unet 7.7 million that increases to 9 million parameters in the Unet++ model and 11.71 million in the NUMSnet model.

The NUMSnet model has two key hyper-parameters. First, the relative location of the training scans in the 3D volume stack impacts the training phase. Since layer information is transmitted to the subsequent ordered scans, selecting training scans that contain the ROIs in several subsequent scans is important. We analyze this sensitivity to training data location in a 3D stack by varying the location of the reference training frame from the beginning to the middle of the stack followed by selecting the subsequent or randomly selected frames in order. For example, this ensures that in the Heart-CT stacks, if an aortic region is detected for the first time in a scan, the ROI first increases and then decreases in size as training progresses. The second hyper-parameter for the NUMSnet model is the number of up-sampling layer features that can be transmitted across scans. If all the 10 up-sampled layer features from layers [X(1,2), X(1,3), X(1,4), X(1,5), X(2,2), X(2,3), X(2,4), X(3,2), X(3,3), X(4,2)] from Fig. \ref{fig:model} are transmitted to the subsequent scans, this would incur higher computational complexity (14.5 million trainable parameters). We analyze the segmentation performance using this NUMSnet model variant (called NUMS-all), where outcomes of 10 up-sampling layers are transmitted. The primary reason for transmitting only up-sampling layers is that up-sampling generates image feature expansion based on pixel neighborhood estimates. Thus, added information during the up-sampling process further aids the foreground versus background decision making process per image plane.

\section{Experiments and Results}\label{exp}
In this work, we analyze the performance of Unet model and variants for multi-class semantic segmentation on volumetric scans using only 10\% of the annotated data for training. To analyze the importance of nested layer propagation across subsequent images, we perform four sets of experiments. First, we comparatively analyze the segmentation performance per ROI for the NUMSnet when compared to Unet \cite{unet} model and its variants\cite{unetp} for the Lung-CT image stacks. Second, we analyze the sensitivity of the NUMSnet model on the relative position and selection of training data for random ordered sampling versus sequential sampling from the beginning or middle of the volumetric stack. Third, we analyze the semantic segmentation performance of the NUMSnet model when only nested layer features are transmitted versus when all up-sampling layer features are transmitted (NUMS-all). Fourth, we asses the semantic segmentation capability of NUMSnet in comparison with Unet variants for transfer learning of weights and biases from segmenting 3 ROIs (in Lung-CT stacks) to segmenting 7 ROIs (in heart-CT stacks).

\subsection{Multi-class Segmentation Performance of Unet Variants}
For any multi-class semantic segmentation model, it is important to assess the computational complexity introduced by additional layers in terms of the number of trainable parameters jointly with the semantic segmentation performances. Table \ref{tab:params} shows the variations in the number of trainable and non-trainable parameters for all the Unet variants analyzed in this work. Here, we find that Unet is the fastest model while NUMS-all has almost twice the number of trainable parameters when compared to Unet. Also, the NUMSnet model is preferable to NUMS-all with regards to computational complexity  as it has lesser chances of overfitting \cite{params}.
\begin{table}[ht!] 
\caption{Variations in the number of parameters in Unet model variants.}
\begin{tabular}{cccc}\hline
Model&Total params&Trainable params&Non-trainable params\\ \hline
Unet&7,767,523& 7,763,555& 3,968\\
wUnet&9,290,998& 9,286,658& 4,340\\
Unet++&9,045,507& 9,043,587& 1,920\\
NUMSnet&11,713,943&11,711,843&2100\\
NUMS-all&14,526,368&14,524,268& 2,100\\ \hline
\label{tab:params}
\end{tabular}
\end{table}

\begin{figure*}[ht!]
    \centering
    \includegraphics[width=5.0in]{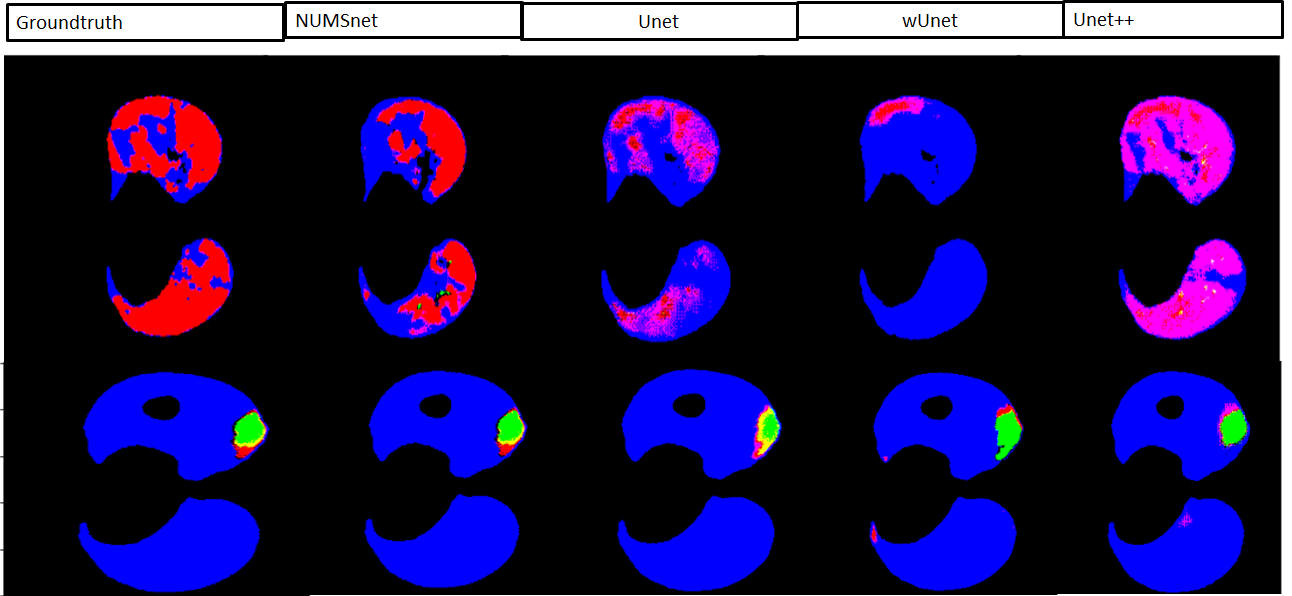}
    \caption{Example of Lung CT segmentation by the Unet variant models. Row 1 represents the poor segmentation results. Row 2 represent good segmentation results since major ROI is the Lung. The color coding is as follows. Blue: Lung regions, Red: GGO regions, Green: Consolidation regions, Magenta: Over detection of Consolidation regions.}
    \label{fig:lung1}
\end{figure*}

Next, we analyze the multi-class semantic segmentation performances of the NUMSnet and Unet model variants. In Table \ref{tab:perf1} the averaged semantic segmentations across 5 random ordered training dataset selections of the 3 regions of interest in Lung-med dataset are presented. Here, we observe that the performance of Lung segmentation is the best and similar across all the Unet variants, with a Dice score ranging between 83-94\%. This is intuitive since the lung is the largest region that is annotated in most images. The Unet and variant models preferentially extract this ROI with minimal training data. However, for segmentation of opacity (GGO) and consolidation (Con) regions, the performance of the NUMSnet model has the highest $Re$ and average Dice scores are 3-9\% better than the Unet++ model. Some examples of Unet and variant model segmentations are shown in Fig \ref{fig:lung1}. Here, we observe that for small as well as large ROIs, the NUMSnet has better segmentation resolution when compared to all other Unet variants. 

For all the Unet variants under analysis the number of epochs is 60 and the optimal loss function is the BDL with Dice coefficient as the reported metric. We observe poor convergence with DL loss function since the large lung regions get weighted more by the DL, thereby resulting in high accuracy of Lung segmentation but poor performances for the GGO and consolidation segmentations.
\begin{table}[ht!] 
\caption{Comparative Performances between the Unet and variant models on the Lung-med stack averaged over 5 runs. Best values are highlighted.}
\begin{tabular}{ccccc}\hline
Task&$Pr$&$Re$&$IoU$&$Dice$\\ \hline
NUMSnet, Con&63.83&	{\bf83.40}&	{\bf55.80}&	{\bf55.90}\\
NUMSnet, GGO&87.92&	{\bf91.79}&	{\bf81.81}&	{\bf84.33}\\
NUMSnet,Lung&{\bf99.61}&	{\bf90.59}&	{\bf90.30}&	{\bf94.00}\\\hline
Unet, Con&{\bf 81.43}&	32.32&	28.25&	31.50\\
Unet, GGO&82.89&	70.30&	60.44&	63.18\\
Unet,Lung&99.17&	90.53&	89.95&	93.84\\\hline
wUnet, Con&64.02&	77.81&	52.99&	53.18\\
wUnet, GGO&73.79&	98.55&	72.98&	73.99\\
wUSnet,Lung&82.50&	90.00&	73.59&	82.88\\\hline
Unet++, Con&69.68&	63.21&	44.57&	46.99\\
Unet++, GGO&{\bf88.97}&	87.60&	79.14&	81.89\\
Unet++,Lung&99.58&	89.80&	89.50&	93.50\\\hline
\label{tab:perf1}
\end{tabular}
\end{table}

\begin{figure*}[ht!]
    \centering
    \includegraphics[width=5.0in]{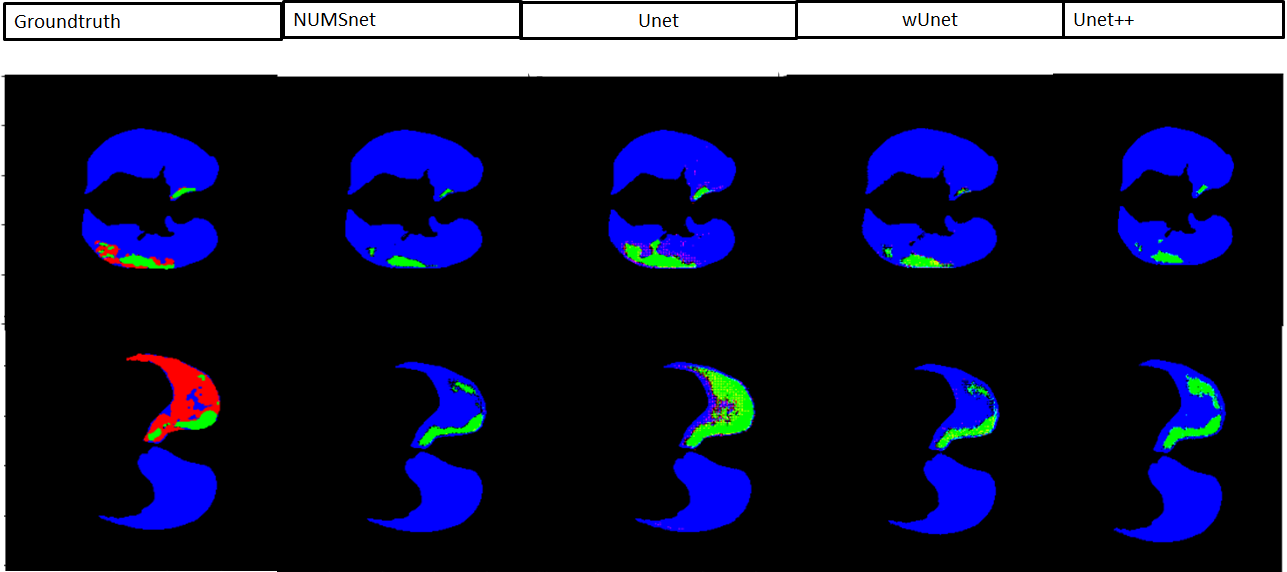}
    \caption{Example of Lung CT segmentation by the Unet variant models. Row 1: Best case detections, Row 2: Worst case detections. The color coding is as follows. Blue: Lung regions, Red: GGO regions, Green: Consolidation regions.}
    \label{fig:lung2}
\end{figure*}

Next, we analyze the segmentation performances on the smaller Lung CT stacks from radiopedia (Lung-rad) in Table \ref{tab:perf2}. Here, we observe that Unet++ has the best performance for segmenting consolidations while NUMSnet has the best performance for segmenting GGO and Lung regions with 3-5\% improved Dice coefficients for the GGO and Lung regions, respectively, over the Unet++ model. Selected good and bad segmentations on this dataset are shown in Fig. \ref{fig:lung2}. Here we observe that the Lung region is well detected by all the Unet model variants, but the Unet misclassifies the GGO as consol (in row 2,  red regions are predicted as green), while the NUMSnet under-predicts the GGO regions. The reason for lesser performance in the Lung-rad stacks when compared to the Lung-med stack is that the number of frames in sequence for training per stack is lesser when compared to the Lung-med stack. Thus, for denser volumetric stacks the NUMSnet has better multi-class segmentation performance when compared to shorter stacks with few images.
\begin{table}[ht!] 
\caption{Averaged performances between the Unet and variant models on 10 Lung-rad CT Stacks across 5 runs. Best values are highlighted.}
\begin{tabular}{ccccc}\hline
Task&$Pr$&$Re$&$IoU$&$Dice$\\ \hline
NUMSnet, Con&64.18&	82.52&	55.45&	55.51\\
NUMSnet, GGO&87.03&	92.59&	{\bf81.74}&	{\bf84.34}\\
NUMSnet,Lung&{\bf99.90}&	{\bf93.03}&	{\bf92.95}&	{\bf95.38}\\\hline
Unet, Con&64.14&70.16&	47.91&	47.93\\
Unet, GGO&72.86&90.70&	70.54&	70.68\\
Unet,Lung&98.47&91.11&	89.74&	93.15\\\hline
wUnet, Con&{\bf67.42}&68.93&	48.02&	49.76\\
wUnet, GGO&{\bf88.21}&90.36&	81.01&	83.59\\
wUSnet,Lung&98.75&92.56&	91.42&	94.18\\\hline
Unet++, Con&64.12&	{\bf99.98}&	{\bf64.5}&	{\bf64.12}\\
Unet++, GGO&82.37&	{\bf95.63}&	79.69&	81.94\\
Unet++,Lung&93.08&	92.40&	86.62&	90.69\\\hline
\label{tab:perf2}
\end{tabular}
\end{table}

\subsection{Sensitivity to Training Data}
In this experiment, we modify the training dataset sequence and observe the segmentation performance variations. We comparatively analyze the performances for three sets of variations in training and test sequences. The first set comprises of training data set that starts from the first scan in the image stack as reference image followed by sequential 10\% of images extracted per stack for training. All remaining images in sequence are considered as test samples, while 1\% images from the test samples are withheld for hyper-parameterization as a validation dataset. This is called the $Initial, Seq$ set. The second set comprises of training images that start from the middle scan per 3D stack. 10\% of the subsequent scans can randomly be selected while maintaining the order of images to generate the training sequence. All remaining images are test data with 1\% images randomly removed as validation dataset. This is called the $Mid, Rand$ set. The third set starts the training images from the middle scan per stack and selects 10\% frames in a sequence as training data. All remaining images are test data with 1\% images separated for validation tasks. This is called the $Mid, Seq$ set. The variations in multi-class semantic segmentations for the Lung-med and Lung-rad scans for all these three training/test stacks is shown in Table \ref{tab:forback}. 
\begin{table}[ht!] 
\caption{Comparative performance of NUMSnet on Lung-CT stacks by varying the Training dataset.}
\begin{tabular}{ccccc}\hline
Task&$Pr$&$Re$&$IoU$&$Dice$\\ \hline
Data:& Lung-med&&&\\
Initial, Seq, Con&82.5& 35.05& 26.1& 29.34\\
Initial, Seq, GGO&85.78& 69.13& 59.30& 62.21\\
Initial, Seq, Lung&88.85& 93.45& 82.98& 89.90\\\hline
Mid, Rand, Con&60.38& 96.52& 57.97& 57.97\\
Mid, Rand, GGO&70.15& 99.46& 69.75& 69.75\\
Mid Rand, Lung&99.27& 89.19& 88.62& 92.94\\\hline
Mid, Seq, Con&60.37& 97.32& 58.91& 58.91\\
Mid, Seq, GGO&70.15&93.17& 68.01& 68.01\\
Mid, Seq, Lung&98.73&89.28& 88.27& 92.59\\ \hline \hline
Data:& 10 Lung-rad &Stacks&&\\ \hline
Initial, Seq, Con&87.74& 44.24& 38.94& 43.13\\ 
Initial, Seq, GGO&92.23& 75.69& 72.46& 75.19\\
Initial, Seq, Lung&95.44& 96.74& 92.87& 95.79\\ \hline
Mid, Rand, Con&62.05& 99.1& 60.91&  60.91\\
Mid, Rand, GGO&72.22&99.0 &70.22&  70.22\\
Mid Rand, Lung&99.79&  91.74&  91.6 &  94.57\\ \hline
Mid, Seq, Con&59.15& 98.51& 59.49&  59.76\\
Mid, Seq, GGO&82.22&99.0 &80.22&  80.22\\
Mid, Seq, Lung&99.0&  90.74&  90.6 &  93.8\\ \hline
\label{tab:forback}
\end{tabular}
\end{table}

Here, we observe that the segmentation performances for the $Initial, Seq$ train/test stack is consistently worse than the training sets that begin at the middle of each volume stack. This is intuitive since the initial layers often contain no annotations or minimal ROIs, being a precursor to the intended ROIs. Thus, using the $Initial, Seq$ training dataset, the NUMSnet model does not learn enough to discern the small ROIs in this stack. Also, we observe that the performances of $Mid, Rand$ and $Mid, Seq$ training stacks are similar for the Lung-med stack. Besides, we observe a 10\% improvement in $Pr$ and $D$ for $Mid, Seq$ over $Mid, Rand$ for GGO segmentations only. Thus, selecting training images in the middle of 3D stacks with random ordered selection is important for training a multi-class NUMSnet model. 

\subsection{Performance Analysis for NUMSnet variants}
In the third experiment we analyze the number of up-sampling layers that should be propagated to subsequent training scans for optimal multi-class segmentation tasks per volume. In Table \ref{tab:weights}, we analyze the segmentation performances of NUMS-all for Lung-CT stacks, where all 10 up-sampling layers are transmitted. Comparing the Dice scores for the Lung-med stack for NUMS-all with those of NUMSnet in Table \ref{tab:perf1}, we observe that NUMS-all improves segmentation $Pr$ but the overall segmentation performances for GGO, Con and Lung are similar. However, for the Lung-rad stacks, comparing Table \ref{tab:weights} and Table \ref{tab:perf2}, we observe an 8\% improvement in consolidations segmentation using NUMS-all. However, give that NUMS-all has higher computational complexity without significant improvement in overall segmentation performances, the NUMSnet model can be considered superior to NUMS-all while training with limited images.
\begin{table}[ht!] 
\caption{Performances for Lung-CT segmentations with NUMS-all model averaged across 5 runs.}
\begin{tabular}{ccccc}\hline
Data& Lung-med&&&\\ \hline
Task&$Pr$&$Re$&$IoU$&$Dice$\\ \hline
NUMS-all, Con&66.81& 72.63& 53.08& 54.86\\
NUMS-all, GGO&83.11& 91.06& 78.09& 81.02\\
NUMS-all, Lung&99.67& 90.93& 90.74& 94.64\\\hline
Data& 10 Lung-rad&Stacks&&\\ \hline
NUMS-all, Con&64.14& 96.04& 63.05& 63.06\\
NUMS-all, GGO&86.97& 92.34& 81.82& 84.34\\
NUMS-all, Lung&99.63& 92.89& 92.56& 95.1\\\hline
\label{tab:weights}
\end{tabular}
\end{table}

\subsection{Transfer Learning for Heart-CT Images}
\begin{figure*}[ht!]
    \centering
    \includegraphics[width=5.0in]{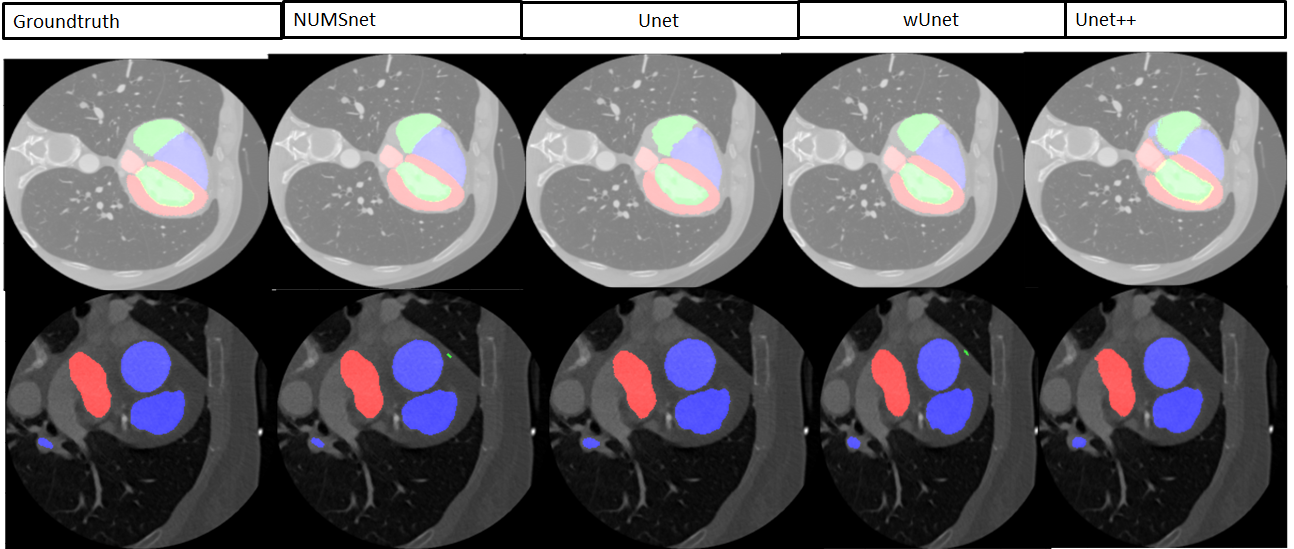}
    \caption{Examples of Heart CT segmentation by the Unet variant models. Row 1: Good segmentations. Row 2: Average segmentations. For Row 2 we observe the variations in the small ROI on the left corner of the image across the Unet variants.}
    \label{fig:heart}
\end{figure*}

Finally, we analyze the transfer learning capabilities of pre-trained Unet and variant models from Lung-CT to the Heart-CT stacks. The trained models from the Lung-med image stack are saved and all layers before the final layer are unfrozen, to be retrained on the Heart-CT dataset. The only difference between the Unet and variant models between the Lung-CT and the Heart-CT image sets is the final number of classes in the last layer X(1,5). Re-using the weights and biases of all other layers provides a warm start to the model and aids faster convergence of the loss functions while training with randomly selected ordered training images. For this experiment, the performance of each Unet variant to segment regions with label pixel values [205, 420, 500, 550, 600, 820, 850] are represented by the model name and [pix$_{205}$, pix$_{420}$, pix$_{500}$, pix$_{550}$, pix$_{600}$, pix$_{820}$, pix$_{850}$], respectively, in Table \ref{tab:heart}.
\begin{table}[ht!] 
\caption{Averaged performances between the Unet and variant models on 10 Heart-CT stacks across 5 runs. Best values are highlighted.}
\begin{tabular}{ccccc}\hline
Task&$Pr$&$Re$&$IoU$&$Dice$\\ \hline
NUMSnet,pix$_{205}$&{\bf96.2}& 78.83& 75.53& 78.01\\
NUMSnet,pix$_{420}$&{\bf96.89}& 86.2& 83.42& 85.04\\
NUMSnet,pix$_{500}$&94.84& {\bf98.16}& {\bf93.29}& {\bf95}\\
NUMSnet,pix$_{550}$&{\bf96.61}& {\bf86.23}& {\bf83.4}&{\bf 85.8}\\
NUMSnet,pix$_{600}$&{\bf94.95}& 80.26& {\bf76.03}& {\bf79.28}\\
NUMSnet,pix$_{820}$&{\bf98.42}& {\bf96.84}& {\bf95.55}& {\bf96.88}\\
NUMSnet,pix$_{850}$&{\bf90.41}& 81.55& {\bf73.03}& {\bf75.05}\\\hline
Unet,pix$_{205}$&95.01& {\bf79.17}& 75.48& {\bf79.3}\\
Unet,pix$_{420}$&95.54& 88.39& 85.31& {\bf87.78}\\
Unet,pix$_{500}$&94.8& 95.08& 91.06& 93.34\\
Unet,pix$_{550}$&92.27& 82.5& 76.44& 80.49\\
Unet,pix$_{600}$&90.09& {\bf83.1}& 74.84& 79.23\\
Unet,pix$_{820}$&97.53& 80.95& 79.43& 81.93\\
Unet,pix$_{850}$&88.1& 46.81& 44.01& 46.81\\\hline
wUnet,pix$_{205}$&95.93& 75.18& 72.37& 76.02\\
wUnet,pix$_{420}$&94.64& {\bf91.75}& 87.37& 89.96\\
wUnet,pix$_{500}$&94.42& 95.05& 90.73& 93.11\\
wUnet,pix$_{550}$&89.27& 64.52& 57.46& 61.54\\
wUnet,pix$_{600}$&90.93& 80.84& 72.93& 77.22\\
wUnet,pix$_{820}$&95.3& 88.77& 84.91& 87.99\\
wUnet,pix$_{850}$&84.37& 69.65& 60.09& 62.74\\\hline
Unet++,pix$_{205}$&96.11&  67.93&  65.01&  68.82\\
Unet++,pix$_{420}$&94.69&  88.92&  84.59&  86.91\\
Unet++,pix$_{500}$&{\bf97.06}&  92.21&  89.93&  92.46\\
Unet++,pix$_{550}$&88.46&  73.42&  63.44&  67.36\\
Unet++,pix$_{600}$&94.21&  73.17&  69.7&  73.09\\
Unet++,pix$_{820}$&96.07&  88.15&  85.06&  86.96\\
Unet++,pix$_{850}$&65.06& {\bf99.95}&  65.07&  65.07\\\hline
\label{tab:heart}
\end{tabular}
\end{table}
Here, we observe that NUMSnet has superior segmentation performances for the smaller ROIs with pixel values [500, 550, 600, 820, 850], respectively, with 2-10\% improvements in Dice scores for these regions over the Unet++ model. Thus, the NUMSnet model aids transfer learning across anatomical image stacks, across label types and yields higher precision for smaller ROIs. Some examples of good and average segmentation using the Unet model variants on the Heart-CT stack are shown in Fig. \ref{fig:heart}. Here, we observe significant variations for smaller ROIs across the Unet model variants.

\section{Conclusion}\label{conclusion}
In this work we present a novel NUMSnet model that is a variation to the Unet++ model \cite{unetp} specifically for multi-class semantic segmentation in 3D medical image stacks using only 10\% of the images per stack selected randomly in an ordered manner around the central scan of the 3D stacks. The novelty of this model lies in the temporal transmission of spatial pixel and neighborhood feature information across scans through the nested layers. The proposed model achieves 3-9\% improvements in Dice scores over Unet++ and other Unet model variants for segmenting 3-7 ROIs per volumetric stack.

The comparative performance of the NUMSnet model with existing works that train deep learning models on larger training datasets is shown in Table \ref{tab:comp}. Here, we observe that the proposed NUMSnet model achieves comparable to improved semantic segmentation performances across a variety of anatomical CT image stacks with only a fraction of the training set of images. This demonstrates the importance of nested layer transmission for enhanced boundary segmentations especially for relatively smaller ROIs. For the Lung-CT stacks, the work in Voulodimos et. al. \cite{covidctlung} introduced few-shot method using a Unet backbone for GGO segmentation only and while this method achieved high precision and accuracy, it had low recall and Dice coefficients. Also, for the same dataset, the work on Saood et. al. \cite{unetsegnet} used a small fraction of images for training, and achieved better binary segmentation performances than multi-class segmentation performances. It is noteworthy that no existing works have bench-marked segmentation performances for the Lung-rad image stacks. For the Heart-CT stacks, most works looked at training on 20 CT stacks and testing on another 20 stacks for high precision of segmentation per ROI. In this work, we apply a pre-trained model on Lung-CT and fine tune it on 4.6\% of all Heart-CT images to obtain similar segmentation performances.
\begin{table}[ht!]
\caption{Comparative performance of NUMSnet with respect to Existing works.}
\begin{tabular}{ccccc}\hline
Method&Data&\#Training Images&Metrics\\ \hline
Saood et. al. \cite{unetsegnet}&Lung-med&72& $D_i=$[22.5-60]\%\\
Voulodimos \cite{covidctlung}&Lung-med&418&$D_i=$[65-85] \%(GGO)\\
Roychowdhury \cite{qunet}&Lung-med&40&$D_i=$64\% (GGO)\\
NUMSnet (Ours)&Lung-med&82&$D_i=$[56-94\%]\\\hline
Wang et. al \cite{unetp2}&Heart CT&7831&$D_i=$[64.82-90.44\%]\\
Payer et. al. \cite{payer}&Heart CT& 7831&$D_i=$[73.7-88\%]\\
Ye et. al. \cite{heartct}&Heart CT&7831&$D_i=$[86-96\%]\\
NUMSnet (Ours)&Heart CT&363&$D_i=$[75-97\%]\\
\label{tab:comp}
\end{tabular}
\end{table}

Additionally, in this work, we analyze a variety of sampling methods to optimally select the minimal 10\% training set. We conclude that random selection of ordered scans is the optimal mechanism to select a minimal training set. Further, we analyzed the optimal number of up-sampling layers that should be transmitted for best semantic segmentation performances. Here, we conclude that the nested layers from a Unet++ model are significant for transmission, while adding additional up-sampling layers for transmission increases the overall computational complexity of the NUMSnet model while not significantly contributing to segmentation performances for sparse training image sets.

Finally, we assess the transfer learning capabilities for the NUMSnet model that is pre-trained on Lung-CT stacks and fine-tuned on Heart-CT images. We conclude that the NUMSnet model aids transfer learning for similar medical image modalities even if the number of classes and ROIs change significantly. This aligns with the works in \cite{transfer} \cite{transfer2} that demonstrate pre-trained models from one medical image modality to scale to other medical image stacks. Future work can be directed towards extending the NUMSnet model to additional medical image modalities such as X-rays, OCT and MRI stacks.

\section{Discussion}\label{disc}
One key limiting condition for semantic segmentation using Unet model and its variants is when scans include written text on them. These irregularities can interfere with the segmentation of the outermost ROIs. In such situations, an overall mask can be generated centered around all the ROI regions and superimposed on the original image before passing it to the Unet and variant models, thus eliminating the written text region for enhanced classification performance. Another alternative for reliable end-to-end segmentation in these cases, if enough annotated images are available, is to train two Unet or variant models to first detect the foreground region in the first Unet variant model followed by segmenting the ROIs in the second Unet model as shown in \cite{unetp2}.

It is noteworthy that single stage Unet model and variants are easily trainable with few annotated images and they typically do not overfit. However, for high resolution images such as whole slide images (WSI), where the dimensions of the medical images are a few thousand pixels per side, resizing such images to smaller dimensions to fit a Unet model or its variant may result in poor segmentation results \cite{wsi}. In such scenarios, splitting the images to smaller patches followed by training Unet model and variants can improve segmentation performances such as shown in \cite{covidctlung}.

One key consideration for multi-class segmentations using Unet variant models is the disparity between the ROI sizes that can significantly impact the training stages when only few annotated training images are available. For example, in the Lung-CT image stacks, the lung regions are relatively larger than the GGO and consolidation areas, because of which using few training images and Dice coefficient loss over hundreds of epochs can bias the model to segment the Lung region only. This occurs since the relative variation in pixel neighborhoods for larger ROIs is smaller than the pixel neighborhoods in smaller ROIs. In such situations, it is crucial to ensure that more training images are selected that have the smaller ROIs annotated and the Unet variant models are run for about 40-60 epochs with region sensitive loss functions.

\begin{figure*}[ht!]
    \centering
    \includegraphics[width=5.5in, height=8in]{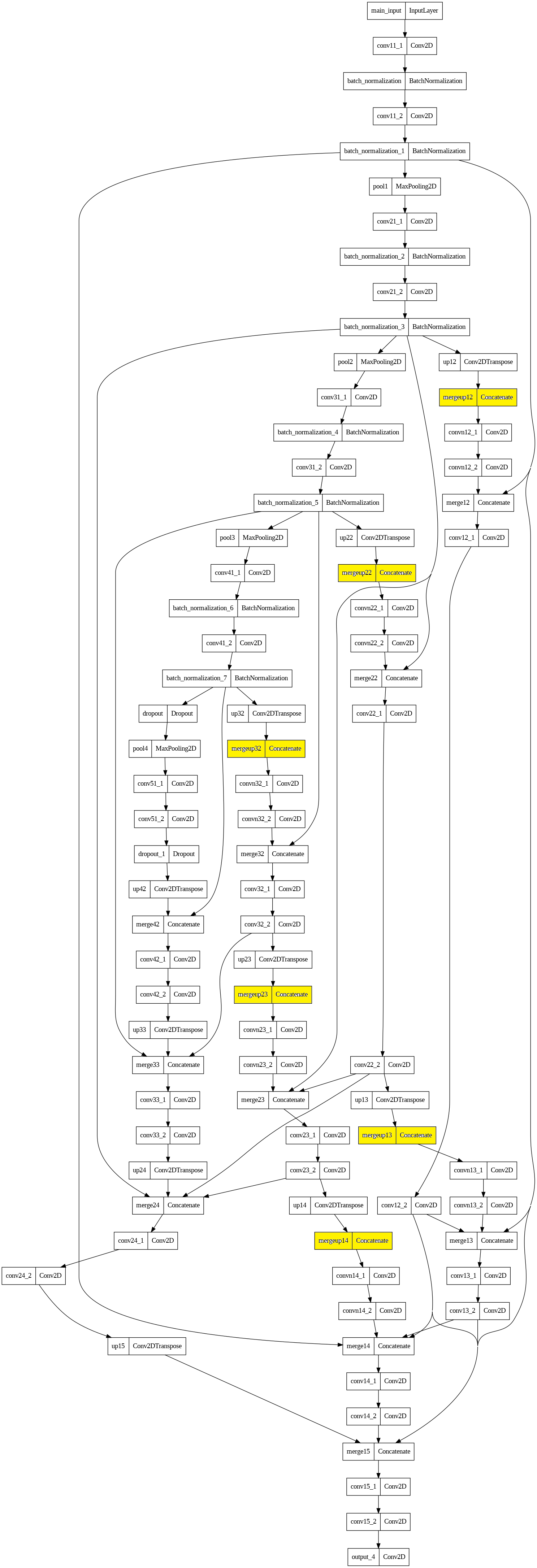}
    \caption{The proposed NUMSnet model. The layers highlighted in yellow are the new concatenation layers introduced by NUMSnet. All the other layers are from the Unet++ model.}
    \label{fig:numsnet}
\end{figure*}

\begin{figure*}[ht!]
    \centering
    \includegraphics[width=5.5in,height=8in]{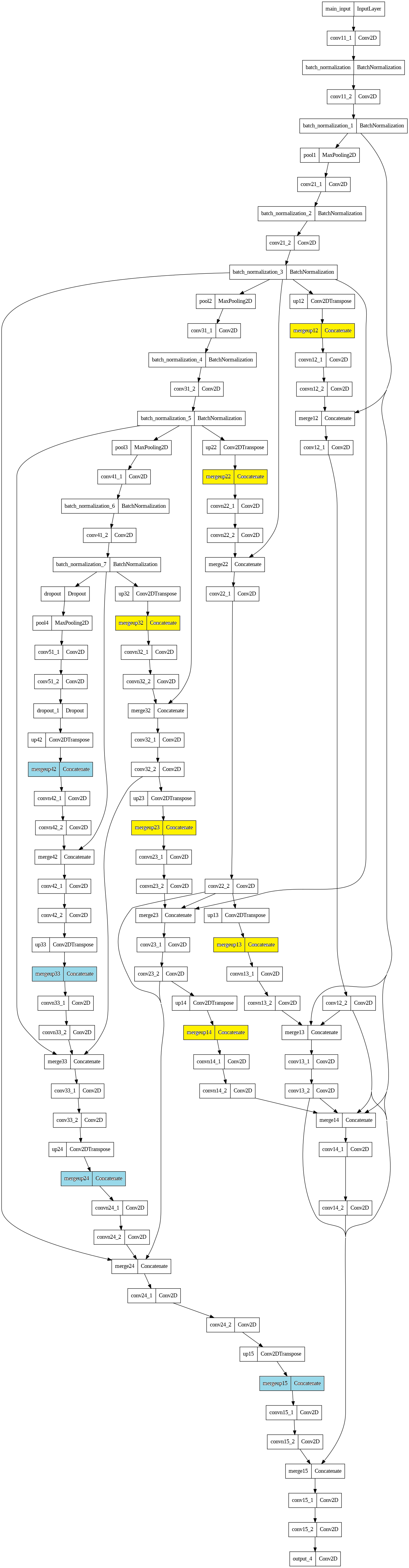}
    \caption{The NUMS-all model with all up-sampled layer features transmitted. The layers highlighted in yellow are the new NUMSnet concatenation layers. The layers highlighted in blue are the additional up-sampled layers in NUMS-all model. All remaining layers are from the Unet++ model.}
    \label{fig:numsneta}
\end{figure*}

Finally, for transfer learning applications, full image network weights transfer better when compared to Unet model variants trained on image patches such as in \cite{covidctlung}. Future efforts can be directed towards transfer learning capabilities of the proposed NUMSnet model on WSI and patch image sets.
\appendix
The proposed NUMsnet model layers and interconnections are shown in Fig. \ref{fig:numsnet}. The layer interconnections from the NUMS-all model are shown in Fig. \ref{fig:numsneta}.

\bibliographystyle{IEEEtran}
\bibliography{main}

%


\end{document}